\documentclass[final,5p,times,twocolumn]{elsarticle}

\usepackage{amssymb}

\usepackage{graphics} 
\usepackage{epsfig} 
\usepackage{amsmath} 
\usepackage{amssymb}  
\usepackage{mathtools} 
\usepackage{bm} 
 \usepackage{multirow}
 \usepackage[linesnumbered,ruled,vlined,Algorithm ]{algorithm }
\usepackage{algpseudocode}
\usepackage{hyperref}
\usepackage{array}
\usepackage[font=small,skip=3pt]{caption}
\usepackage{subcaption}
\usepackage{enumerate}
\usepackage[english]{babel} 
\usepackage{amsfonts,amsthm,bm} 
 \usepackage{cleveref}

\usepackage{amsmath}
\usepackage{graphicx}
\usepackage{wrapfig }
\usepackage{multirow}
 \usepackage[edges]{forest}
\usepackage{tcolorbox}
\usepackage{tabularx,colortbl}
\usepackage{adjustbox}
\usepackage{colortbl}
\usepackage{hhline}
\usepackage{breqn}

\def\middlebreak {\nulldelimiterspace0pt
\right.\allowbreak\mskip 0mu plus .5mu \nulldelimiterspace0pt\left.}%
 
\journal{XXXXX}

\begin{document}

\begin{frontmatter}

\title{A Dynamic Heterogeneous Team-based Non-iterative Approach for Online Pick-up and Just-In-Time Delivery Problems}

\author[inst1]{Shridhar Velhal}
\author[inst1]{Srikrishna B R}
\author[inst2]{Mukunda Bharatheesha}
\author[inst1,]{Suresh Sundaram}

\affiliation[inst1]{organization={Department of Aerospace Engineering,},
            addressline={Indian Institute of Science, Bangalore},
            country={India}}
\affiliation[inst2]{organization={Robert Bosch Centre for Cyber-Physical Systems,},
            addressline={Indian Institute of Science, Bangalore},
            country={India}}

\begin{abstract}
 This paper presents a non-iterative approach for finding the assignment of heterogeneous robots to efficiently execute online Pickup and Just-In-Time Delivery (PJITD) tasks with optimal resource utilization. 
The PJITD  assignments problem is formulated as a spatio-temporal multi-task assignment (STMTA) problem. The physical constraints on the map and vehicle dynamics are incorporated in the cost formulation. The linear sum assignment problem is formulated for the heterogeneous STMTA problem. The recently proposed Dynamic Resource Allocation with Multi-task assignments (DREAM) approach has been modified to solve the heterogeneous PJITD problem. At the start, it computes the minimum number of robots required (with their types) to execute given heterogeneous PJITD tasks. These required robots are added to the team to guarantee the feasibility of all PJITD tasks. Then robots in an updated team are assigned to execute the PJITD tasks while minimizing the total cost for the team to execute all PJITD tasks. The performance of the proposed non-iterative approach has been validated using high-fidelity software-in-loop simulations and hardware experiments. The simulations and experimental results clearly indicate that the proposed approach is scalable and provides optimal resource utilization.
\end{abstract}



\end{frontmatter}

\section{Introduction}
With growing technology, robots have been used in various industrial applications. Multi-robot systems provide a distributed, reliable, and scalable approach for handling various operations. With the help of developments in IoT and Industry 4.0 technologies, just-in-time \citep{monden2011toyota,emde2012optimally} approaches are used in the automation industry to manage storage and inventories optimally. 
Warehouses are the critical connection hub in the supply chain of the e-commerce industry, and warehouse automation is becoming very important \citep{boysen2019warehousing,wurman2008coordinating,fischer2020design,da2021robotic}.  
Customers demand quick and on-time delivery of items, and time is becoming a crucial aspect of the e-commerce industry. The time sensitivity of delivery tasks increases for perishable items such as food and beverages. Due to the ever-persistent competition in e-commerce, even-for non-perishable items, the on-time delivery of items is a game-changing factor. Warehouse management and e-commerce are a few of the important applications that require online solutions and whose efficacy can be improved with the use of JIT tasks. 

A typical warehouse has many objects that need picking and placing between various locations, which is currently done by autonomous robots. If items are delivered at an exact time, the subsequent processes can start immediately, improving the efficiency of operations. Also, it will reduce/eliminate the need for local storage space. The packaging of different items for an order is one example where all items must be at the packaging counter at the desired time. Local storage is not required if all items come to the packaging counter at desired times; it also helps improve efficacy by reducing redundant pick and place operations.
Just-in-time (JIT) management strategy implemented in manufacturing and automobile industries to align raw-material orders from suppliers directly with production schedules. A major concern in JIT approach is the potential disruptions in the supply chain. In this paper, we propose use of robots for pick-up and just-in-time delivery tasks in warehouse operations and get the benefits of the JIT approach with a robust supply chain maintained with robots. 
\begin{figure}[t!]
    \centering
    \includegraphics[width=\linewidth]{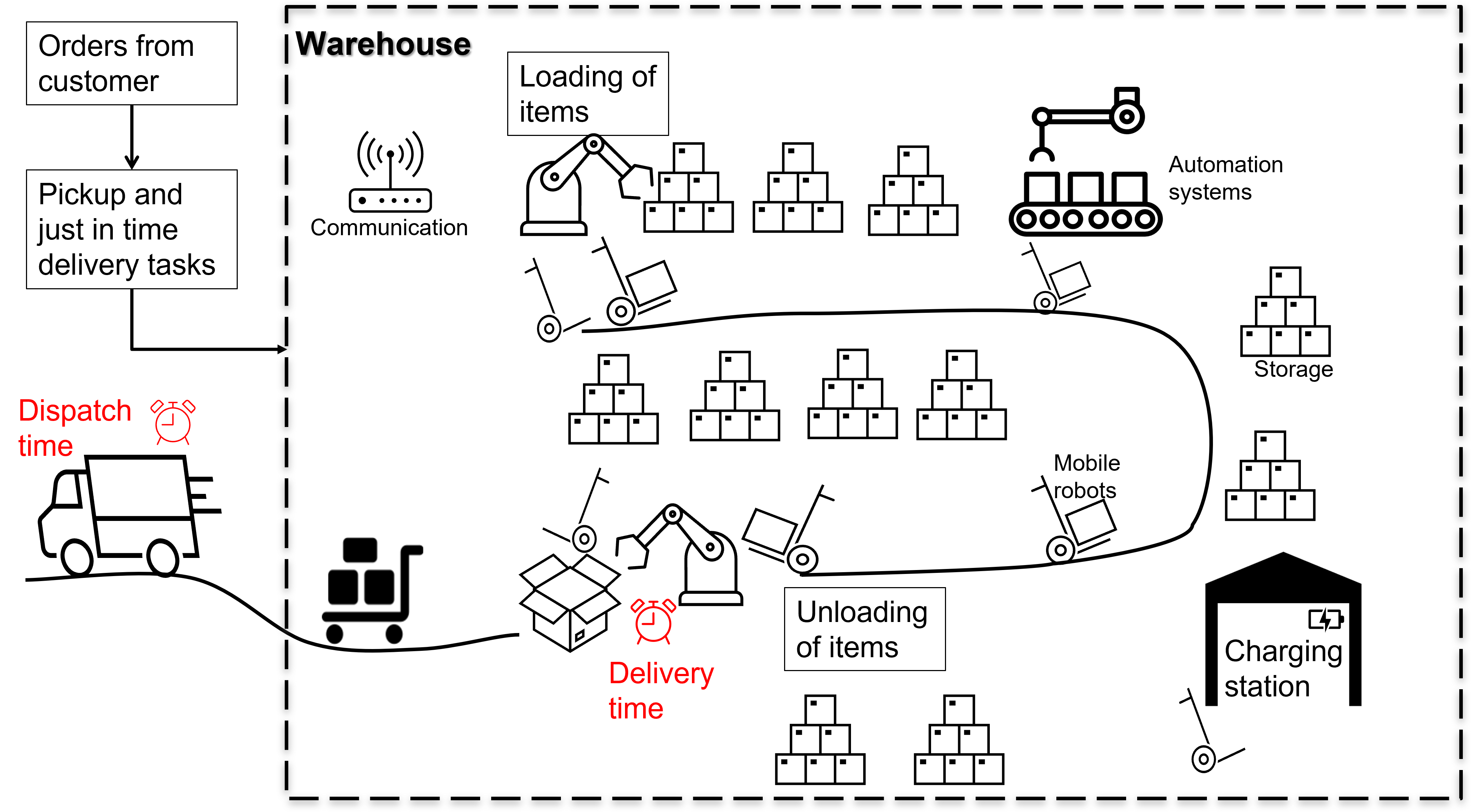}
    \caption{Typical Warehouse and its operations }
    \label{fig:warehouse_automation}
\end{figure}

A multi-robot pick-up and delivery problem have been approached via distributed resource allocation in \citep{camisa2021multi}. The cost function minimizes the total distance traveled by robots while executing pick-up and delivery tasks.  In \citep{chen2021integrated}, an integrated approach for task assignment and path planning for capacity-constrained multi-agent pick-up and delivery problems have been presented. This approach also handles multiple packets carrying during transport. The marginal-cost-based and regret-based marginal-cost-based algorithms minimize the total travel delay while avoiding collisions. The multi-task allocation problem for the final-mile delivery using drones has been solved by \citep{ham2019drone}, where a drone has to pick and deliver an item one by one. 
In \citep{bolu2021adaptive}, the adaptive task allocation in warehouse operations has been presented to handle system dynamics such as the location of tasks, number of robots, replenishment of new stock, and battery of robots.
In \citep{bai2019efficient}, the complexity of this combined pick-up and delivery has been reduced by considering them as separate tasks and putting precedence constraints such that the same robot should pick up the items before delivery for a last-mile delivery problem. But this increases the number of tasks to double. 

The detailed review of the tasks assignments and scheduling with different temporal constraints has been presented in \citep{nunes2017taxonomy}. 
Two critical problems,  deadline and time-window problems\citep{vu2018solving,jiang2020traveling,jiang2021order,gao2020approximation,semiz2020solving}, are well-studied in the literature for warehouse problems.
The deadline  tasks require local storage to keep items before they are used for the subsequent process (which will start only after the deadline). The time-window problems require the items to be delivered within a time window and the local storage is available only for that time window.  
The traveling salesman problem with time windows \citep{vu2020dynamic} provides the mathematical framework for the time-constrained TSP. 
The solution approach considers the pick-up and delivery as separate tasks and adds the constraint on a robot that picks the item should deliver the item, and only delivery is allowed only after the pick-up. This increases the constraints and dimensions of the optimization problem.
In \citep{budak2020evaluation}, the effects of the size of the time window were studied, and it is observed that decreasing the size of the time window increases customer satisfaction and decreases computation time, but it increases the tour duration.

All the aforementioned works in warehouse automation do not consider the JIT tasks and assume the feasibility of tasks for a given team, so they assume only a fixed-sized team of robots. Hence there is a strong need to develop an algorithm to handle the heterogeneous JIT tasks with a dynamic-sized team of heterogeneous robots and compute the feasible solution by utilizing the minimum resources (robots).

In this paper, we propose heterogeneous resource allocation approach for the online pick-up and just-in-time delivery problem with heterogeneous robots in warehouse management. 
The cost function defined in \citep{velhal2022dynamic} has been modified for heterogeneous robots to execute the heterogeneous tasks. 
The cost modification also considers the total distances traveled by robots for pick-up and delivery, the loading time (at pick-up) and unloading time (at delivery) required by robots.
The proposed heterogeneous resource allocation approach for PJITD tasks provides the non-iterative solution that computes the optimal trajectories for a dynamic-sized team of robots to execute given heterogeneous spatio-temporal tasks. In the first step, the number of robots required to execute the given heterogeneous spatio-temporal task is computed. Those many robots from different skill sets are added to the active team of robots, and finally, feasible assignments for an updated team of heterogeneous robots are computed. This way, in at most two steps, one can compute the optimal assignments to execute the given heterogeneous spatio-temporal tasks with minimum resources (robots).
From the solution of heterogeneous spatio-temporal multi-task assignment (STMTA), the trajectories of active robots are computed using the trajectory generation algorithm, following which a team of robots will execute all the given tasks.
The working of the DREAM algorithm for PJITD tasks is presented in both simulation and hardware experiments. The high-fidelity simulations are carried out in a ROS2-Gazebo environment. The lab-scale hardware experiments are conducted to illustrate the working of the proposed heterogeneous resource allocation approach for PJITD problems.

The rest of the paper is organized as follows;  Section \ref{sec:related_work}  provides the related works. Section \ref{sec:PJITD_problem_formulation} presents the mathematical problem formulation for the online pick-up and just-in-time delivery task assignment problem. Section \ref{sec:STMTA-for_PJITD} presents the heterogeneous resource allocation approach for computing the feasible task assignments. The working of the proposed approach is shown in Section \ref{sec:result}. The paper is concluded in Section \ref{sec:conclusion}

\section{Related works}  \label{sec:related_work}
The proposed work uses the idea of JIT tasks, the spatio-temporal multi-task assignment problem, and a dynamic-sized team of robots to execute given spatio-temporal tasks. Here, we briefly review these related works.
\subsection{Just-in-Time (JIT)}
The just-in-time \citep{monden2011toyota,emde2012optimally} approach demands the tasks should be done on exact time; this helps to manage the inventory and storage optimally. Recently, a new approach named zero-warehousing and smart manufacturing has been presented in \citep{lyu2020towards} in which IoT-based zero-warehousing is proposed to minimize the non-value adding and redundant handling warehouse process and also to minimize the warehousing space. Recently, \citep{nishida2019just,nishida2022dynamic} have presented the just-in-time approach for pick-up and delivery for automated guided vehicles. The cost function has been formulated to minimize the deviations from the desired pick-up and delivery times; hence, it handles the temporal constraints softly. The aforementioned works on JIT are designed for static environments where tasks are known in advance and solution can be computed offline.

\subsection{Spatio-Temporal Multi-Task assignment (STMTA) }
Chopra and Egerstedt \citep{chopra2014heterogeneous,chopra2015spatio} has presented the multi-robot routing problem and demonstrated using the music wall, where the robot reaches different note location and plays musical notes at respective specific exact times.  As spatio-temporal tasks need to be done at the desired times, some minimum number of robots is required. The main issue in the spatio-temporal task assignment is the computation of the minimum number of robots required to execute the given spatio-temporal tasks. 
In \citep{chopra2014heterogeneous,chopra2015spatio}, the required minimum number of robots is computed offline, in an iterative way, for given tasks. This iterative method for computing required a minimum number of robots and was a big huddle for the online use of STMTA. 

\subsection{Dynamic resource allocation approaches for STMTA}
Dynamic REsource Allocation with decentralized Multi-task assignment (DREAM) \citep{velhal2022dynamic} approach has been proposed for the spatio-temporal multi-task assignment problem. It provides the non-iterative solution to compute the required number of homogeneous robots to execute the given spatio-temporal tasks and their assignments to execute the given spatio-temporal tasks. The non-iterative DREAM approach has been implemented to compute the collision-free trajectories for a dynamic-sized team of music-playing robots (i.e., just-in-time tasks with homogeneous robots) in \citep{velhal2022non}.  The DREAM approach is limited to homogeneous agents and considers only simple routing tasks.  PJITD tasks demand a solution for heterogeneous robots, so DREAM is not directly applicable to PJITD tasks.

The warehouse automation requires a non-iterative (online) solution to assign multiple complex tasks (a combination of a few sub-tasks and waiting) to the optimal-sized team of heterogeneous (with different speed and payload carrying capability) robots. The DREAM algorithm provides an online solution but considers simple routing tasks homogeneous robots. Hence there is a need to develop an online implementable algorithm that handles the online, dynamic and heterogeneous complex tasks in a warehouse environment.

\section{Pick-up and Just-In-Time Delivery Tasks} \label{sec:PJITD_problem_formulation}
Typical warehouse operations are shown in Fig. \ref{fig:warehouse_automation}. The main operational objective in the warehouse is to minimize the time of dispatch of items from the warehouse once the order is received.  
The ordered items are dispatched from a warehouse to some local hub near the customer. All items belonging to one local hub need to be collected on priority before the scheduled leaving time of the vehicle transporting items to that hub/customer. 
One can use linear temporal logic approaches \citep{baier2008principles}  to generate sub-tasks in automation; bin packaging algorithms \citep{leao2020irregular} to compute the sequence in which items need to be packed.  As all items in a single package should be packed together, the exact delivery time will help speed up the packaging process. All items from a single package can come together and be directly packed without local storage and time delay. It also eliminates the redundant pick and place operations for local storage. In this way, the efficacy of operations will be improved.

In warehouse operations, a human operator (near pick-up) or an arm system (on each robot) is required to pick up multiple items and transport them. To avoid the complexity of the co-working environment, this paper assumes that one robot can execute one task at a time. i.e., if a robot has picked one item, it has to deliver it before picking another. A robot can plan for future tasks but must complete the first task before starting the next one. 
 
In this paper, we assume that once the order is received from customers, the sub-tasks are defined, and pick-up and just-in-time delivery (PJITD) problems are generated. This paper presents the solution to the PJITD problem while minimizing the robots required and the collective distance traveled by the dynamic-sized team of robots to execute all PJITD tasks on time. 
The PJITD task demands the robots with the desired skill set  to pick up the items and deliver them at a specific location at a specific time; hence this task is also called a  heterogeneous spatio-temporal task. 
One should note that the given JIT/spatio-temporal tasks will require a minimum number of robots to execute the tasks on time. 
The objective of this paper is to compute the assignments of robots to execute online heterogeneous tasks while minimizing the resources (i.e., the number of actively used robots) and the total distance traveled. 
 
First, we define the notations used in the paper. \newline
$\bm{p} = (x,y) \in \mathbb{R}^2 $: location in Cartesian coordinate  \\
$\bm{p}^R $:  robot's location \\
$\bm{p}^P $ :  pick-up location \\
$\bm{p}^D $:   delivery location\\
$ R_i$ : robot number $i$, ($i$ is used for robot ) \\
$\overline{V}_{i}^R $: maximum velocity of $R_i$  \\
$Q_{\ell}  $: quality/skill set  $\ell,  \ \ell \in \mathcal{L} = \{ 1,2,\cdots,n_\ell \} $, \\
$Q(R_i)$ :  set of qualities/skills of robots $R_i$ \\
$\tau^l$ : loading time \\
$\tau^u$ : unloading time \\
$t_j^D$ : delivery time of task $T_j$ \\
$Q(T_j)$ = set of qualities/skills required to  execute the task $T_j$. \\
$ T_j({\bm{p}}_j^P, \tau_j^l, {\bm{p}}_j^D, \tau_j^u, t_j^D ,Q(T_j))  \text{ or }  T_j :  $ $j^{th}$ task  ( indices $j$ and $k$ are used for tasks)\\
$\mu_i$ : sequence of tasks assigned to robot $R_i$  \\
$c_{ij}^{f,Q(R_i)}$ : cost of $R_i$ (with quality $Q(R_i)$) to  execute the  $T_j$ as a first task \\
$c_{kj}^{s,Q_\ell}$ : cost of robot with quality $Q_\ell$   to execute the  task $T_j$ just after the task $T_k$ (subsequent task).\\
$\delta_{ij}^{f,Q(R_i)}$ : decision variable whether robot $Q(R_i)$ execute the task $T_j$ as first task or not.\\
$\delta_{kj}^{s,Q_\ell}$ : decision variable whether robot with quality $Q_\ell$ execute the task $T_j$ just after task $T_k$ or not. 
 
\subsection{Mathematical Formulation}
Consider  a  set of $N$ robots denoted as  $\mathcal{R}$,  $ \mathcal{R} =  \{R_1,R_2,\cdots,R_N\}$. 
The position of robots $R_i$ is denoted as ${\bm{p}}_i^R = (x_i^R , y_i^R)$. 
Typically robots have different finite skills (for example, the weight carrying capacity, size carrying capacity), and they are represented using set,   $lambda_ell$,   
Robots will be assigned to pick-up and delivery tasks. The location of pick-up for task $T_j$ is denoted as  ${\bm{p}}_j^P = (x_j^P, y_j^P)$ and location of delivery for task $T_j$ is denoted as ${\bm{p}}_j^D = (x_j^D, y_j^D)$. 
The charging stations are placed at $S_i = (x_i^{C},y_i^{C})$. A warehouse robot operates in two modes; active mode, when robots are assigned to a task, and rest mode. In rest mode, a robot is either ideal or charging its battery from the charging station. 
 
\subsubsection{ Pickup and just-in-time delivery task }
A pickup and just-in-time delivery (PJITD) task  consists of loading an item (the robot has to stop at the picking station for loading time), traveling to the delivery location, and unloading the item (the robot has to wait for unloading time). This execution has to be completed on the desired delivery time. If a robot starts a task and executes it on the desired delivery time, then the PJITD task is completed. If the robot reaches the delivery location after the desired delivery time, the task execution fails.
\\
Consider a PJITD task ($T_j$), in which a robot has to visit the pick-up location (${\bm{p}}_j^P$), wait for the loading time ($\tau_j^l$) to load items on a robot. After picking items, a robot should reach the desired packing/processing (delivery) counter located at (${\bm{p}}_j^D $), unload the items with unloading time ($\tau_j^u$) on or before the delivery time ($t_j^D $). This task is represented  by $T_j({\bm{p}}_j^P, \tau_j^l, {\bm{p}}_j^D, t_j^D, \tau_j^u,Q(T_j)) $;  in rest of the paper, this  PJITD task is referred as $T_j$.

A task consists of sub-tasks listed below:
\begin{enumerate}[(a)]
\itemsep0em
    \item select the robot with desired skills (i.e. $ Q(T_j) \subseteq  Q(R_i) $ )
    \item robot should reach to the pick-up location ${\bm{p}}_j^P$
    \item wait for  $\tau_j^l$ to load the items 
    \item reach to the delivery/drop location ${\bm{p}}_j^D$
    \item wait till the delivery time $t_j^D$
    \item wait for $\tau_j^u$ to unload the items
\end{enumerate}

The spatial distance traveled in the execution of a task is the distance between the current position of the robot to the pick-up location (${\bm{p}}_j^P $) and then travel to the delivery station at ${\bm{p}}_j^D$. This task has to be executed with delivery time constraints.
 
Once orders are received from a customer, tasks are generated, and the task allocator will assign the task to the robots. The robot has to execute the assigned tasks in sequence. Lets say the task assigned to robot $R_i$ is $\mu_i$, $\mu_i = \{T_a, T_b\}$ then robot starts from its initial position, moves to pick up the items from the location of task $T_a$,  (i.e., ${\bm{p}}_a^P$), then delivers it to the delivery location of task $T_a$ (i.e., ${\bm{p}}_a^D$) on or before the delivery time $t_a^D$.
Now, from the delivery point of task $T_a$, the robot moves to the pick-up location of task $T_b$, picks the task, and delivers it to the delivery location ${\bm{p}}_b^D$. In short, from the previous task's delivery location, robots move toward the next assigned task's pick-up location and execute the tasks in sequence.
 
Let us consider the PJITD tasks available at any given time $t$ be $\left\{T_1\left(\bm{p}_1^T,t_1 \right), \middlebreak   T_2\left(\bm{p}_2^T,t_2\right), \middlebreak  \cdots, \middlebreak T_{M_t}\left(\bm{p}_{M_t}^T,t_{M_t} \right)  \right\}$.  In general, the number of tasks ($M_t$) are more than the number of robots($N$),   $(M_t > N)$. Note that the number of tasks depends on the customer's order, and a new task is added for every new order from the customer. However, for a given time, $t$, the number of tasks is $M_t$. Let us consider, given $M_t$ tasks are feasible with the team of $N$ robots, then these $N$ robots need to plan their trajectories (set of pick and delivery points with respective delivery times) $ \left\{\mu_1,\ \mu_2,\ \cdots,\ \mu_N\ \right\} $, cooperatively such that, collectively robots completes all the PJITD tasks. 
 
The main objective of this PJITD problem is to find the optimal assignment of multiple  tasks to the robots such that all PJITD tasks are executed while minimizing the total distance traveled and optimizing resource utilization. The major challenge in this PJITD task assignment is to compute the minimum number of robots required to execute all the given PJITD tasks. Once the minimum robots are identified, the heterogeneous spatio-temporal multi-task assignment problem is solved to compute the feasible and optimal trajectories for robots. The details of this approach are explained in the next section.
 
\section{Heterogeneous Resource allocation approach for PJITD task assignments problem}\label{sec:STMTA-for_PJITD}
In the previous section, the PJITD task has been discussed. This section provides the algorithm for assigning robots to execute those PJITD tasks. In order to execute the given PJITD tasks, robots need to visit the sequence of locations at specific times. For pick-up of items robot can visit at any time, but for delivery, a robot needs to be at the delivery location at the desired delivery time. Due to the temporal constraint,  some minimum number of robots is required to execute all the given heterogeneous spatio-temporal tasks. Hence, a dynamic-sized team of robots is used. The minimum required number of robots is utilized to execute given PJITD tasks. 
The proposed algorithm computes the minimum number of robots required to execute given PJITD tasks. Next, it computes the optimal trajectories for robots in the updated team such that they execute all given PJITD tasks. 
The robots are assigned tasks to minimize the total cost, and the cost for executing the PJITD task is described in the subsection. 
 
Each robot executes assigned tasks; starting from its current position, it picks up the items from the pick-up point and delivers them at the exact desired delivery time at the delivery location. Next, the robot moves towards the pick-up of the assigned subsequent task. The robot executes its assigned tasks in sequence;  sequence constitutes both the spatial locations and times the sequence is referred to as the \emph{trajectory} of that robot ($\mu$) (as it constitutes both the spatial locations and times). The feasible trajectories of all robots are computed with the actions of a robot from its current position and then the delivery locations of assigned tasks.
 
The binary decision vector $ \bm{\delta}$ has two components, namely the first assigned task ($ \bm{\delta}_i^{f,Q(R_i)} \in \mathbb{R}^{M_t} $) and the second being the subsequently assigned tasks ($ \bm{\delta}_k^{s,Q_\ell} \in \mathbb{R}^{M_t}$). 
The decision variables are given as
$\bm{\delta}_i^{f,Q(R_i)} =   \left[  \delta_{i1}^{f,Q(R_i)},  \middlebreak \delta_{i2}^{f,Q(R_i)}, \middlebreak  \cdots, \middlebreak \delta_{iM_t}^{f,Q(R_i)} \right] $,    $ i= \{1,2,\cdots,N \}$, and 
$\bm{\delta}_k^{s,Q_\ell}   =  \left[  \delta_{k1}^{s,Q_\ell},  \middlebreak \delta_{k2}^{s,Q_\ell}, \middlebreak \cdots, \middlebreak \delta_{kM_t}^{s,Q_\ell} \right]$, 
$k=\left\{ 1,2, \middlebreak \cdots, \middlebreak  M_t - 1  \right\}, \   \ell=\{ 1,2,\cdots,n_\ell \}$.
 
The first decision variable $\delta_{ij}^{f,Q(R_i)} \in \{1,0\}$ denotes whether the robot $R_i$ first executes the task $T_j$ or not. The subsequent decision variables $\delta_{kj}^{s,Q_\ell} \in \{1,0\}$ denote whether the robot $R_i$ will execute the task $T_j$  just after the execution of the task $T_k$ or not.  
The decision variables, $\bm{\delta}^{f,Q(R_i)}$ and $\bm{\delta}^{s,Q_\ell}$ are optimized to minimize the cost (fuel spent), which is based on the distance traveled by a robot to execute all PJITD tasks at their respective delivery times. 

\subsection{Cost Function}
The cost of a task is defined as the distance that needs to be traveled by a robot from its location to the pick-up location and then to the delivery location before the desired delivery time, and the robot has the required skill to execute the task. For a robot with required skills executing its first task from its initial position, the cost of the first task (${\bf{C}}^{f,Q(R_i)}$) is the distance traveled by a robot $R_i$ from its current position to the pick-up location and from the pick-up location to the delivery location on or before the delivery time of PJITD task. Mathematically, 
 \begin{align}
 & d_1(R_i,T_j) = d(   {\bm{p}}_i^R ,  {\bm{p}}_j^P  ) +  d(   {\bm{p}}_j^P ,{\bm{p}}_j^D  )   \\
C^{f,Q(R_i)}_{ij} &= \begin{cases}       d_1(R_i,T_j)      & \text{if } \dfrac{  d_1(R_i,T_j)  }{\overline{V}_{i}^R}  \le  t_j^D -  \tau_j^l   \\  
\kappa    & \text{if } \dfrac{  d_1(R_i,T_j)  }{\overline{V}_{i}^R}  >  t_j^D -  \tau_j^l   \\  
\kappa & \text{if }  Q^R_i \nsubseteq Q^T_j  \\
\end{cases} \label{eq:cost_cf}    \\  
    \text{for } i & \in \mathcal{I} = \{1,2,\cdots,N\} ,\quad   j \in \mathcal{J} = \{1,2,\cdots,M_t \} \qquad   \nonumber 
\end{align}
where $\kappa$ is a large value,  and $d(A,B)$ is the  distance along the shortest feasible path from point A to point B.
 
The cost for executing the subsequent tasks (${\bf{C}}^{s,Q_\ell}$)  by the robot $R_i$ is the distance traveled by the robot from the location of its previous delivery location to reach the location of the current pick-up location and from the current pick-up location to the current delivery location on or before the delivery time of that subsequent task. If  a robot does not have the skill set to execute the task, then the cost is set to $\kappa$. If a task is required to be executed in negative time, then the cost is set to $\infty$.
\begin{align}
 & d_2(T_k,T_j) =  d(   {\bm{p}}_k^D,  {\bm{p}}_j^P ) + d ({\bm{p}}_j^P , {\bm{p}}_j^D )  \\
 C^{s,Q_\ell}_{kj} &= \begin{cases}  d_2(T_k,T_j)  & \text{if } t_{  k,j}^{\ell} \le  t_j^D - ( t_k^D + \tau_k^u  + \tau_j^{l}    )   \\
\kappa  & \text{if }  t_{  k,j}^{\ell} >  \left(  t_j^D - ( t_k^D  +  \tau_k^u + \tau_j^l  )  \right)  >0 \\
\kappa & \text{if }  Q(T_j) \nsubseteq   Q_\ell \\
\infty & \text{if } \left(  t_j^D - ( t_k^D  +  \tau_k^u  + \tau_j^l )  \right)  \le 0 \\
\end{cases}  \label{eq:cost_cs}      \\ 
& \text{for } k \in   \mathcal{K}  = \{ 1,2,\cdots,M_t-1 \}; \ j \in \mathcal{J};\ \ell \in \mathcal{L} \nonumber    
\end{align}
where $ t_{  k,j}^{\ell} $ is the minimum time required by robot $R$ with quality $Q_\ell$ to travel from the location of task $T_k$ to task $T_j$ and it is computed as
\begin{align} 
t_{  k,j}^{\ell}  = \dfrac{ d_2(T_k,T_j)  }{\overline{V}_{\ell} }   
\end{align}
 
\subsection{Optimization Problem} 
A heterogeneous resource allocation algorithm assigns robots to execute multiple PJITD tasks. A robot will execute the tasks in a sequence, and we denote the sequence assigned to robot $R_i$ by $\mu_i$. Here, the problem of computing sequence $\mu_i$ has been converted to compute each move of one robot from one location to another; combining all moves, one can get the sequence of tasks. Each robot computes its sequence such that it executes all PJITD tasks while minimizing the distance traveled. The first decision variable $\delta^{f,Q(R_i)}_{ij}$ is used to denote that either a robot $R_i$ moves from position ${\bm{p}}^R_i$ to the position ${\bm{p}}^P_j$ picks up the task and then deliver it to the delivery station ${\bm{p}}_j^D $ at time $t'_j$ and $t'_j  \le t_j^D$. The subsequent decision variable   $\delta^{s,Q_\ell}_{kj}$ is used to denote that either a robot $R_i$ moves from its previous task's delivery position ${\bm{p}}^D_k$ at the time ($t'_k$)   to the next task's pick-up position   ${\bm{p}}^P_j$ and deliver the item to the delivery station (${\bm{p}}_j^D$) on or before the delivery time  $t^D_j$. 
The integer programming problem is defined as,
\begin{subequations} 
  \addtocounter{equation}{-1}
\begin{align} \label{eq:integer_prog}
\begin{split}
   \quad &   \min_{\delta^{f,Q(R_i)}_{ij} \ \delta^{s,Q_\ell}_{kj}}     \sum_{i\in \mathcal{I} }  \sum_{j  \in  \mathcal{J}    }  C^{f,Q(R_i)}_{ij} \delta^{f,Q(R_i)}_{ij}   \\ 
     &  \qquad  \qquad   \quad  +    \sum_{\ell\in \mathcal{L} } \sum_{k\in \mathcal{K} }    \sum_{j  \in   \mathcal{J}    }  C^{s,Q_\ell}_{kj} \delta^{s,Q_\ell}_{kj}    
\end{split}\\ 
 {\rm s. \ t.} \   
 & \delta^{f,Q(R_i)}_{ij} \in \{0,1\}\qquad \forall (i,j) \in {  \mathcal{I} \times {\mathcal{J}  } } \label{eq:cost_cond_1a} \\
 & \delta^{s,Q_\ell}_{kj} \in \{0,1\}\qquad \forall (\ell, k,j) \in {  \mathcal{L} \times \mathcal{K}   \times {\mathcal{J} } } \label{eq:cost_cond_1b} \\
 & \sum_{i\in \mathcal{I} }  \delta^{f,Q(R_i)}_{ij} +   \sum_{\ell \in \mathcal{L}} \sum_{k \in {\mathcal{K}  } }  \delta^{s,Q_\ell}_{kj} = 1  \quad  \forall j \in  {\mathcal J} \label{eq:cost_cond_2} 
 \\
 &\sum_{j \in {\mathcal{J} }}  \delta^{f,Q(R_i)}_{ij}   \le 1    \quad  \forall  i \in   \mathcal{I}     \label{eq:cost_cond_3}  
 \\
 &   \sum_{ \ell \in \mathcal{L} } \sum_{j \in {\mathcal{J} }}  \delta^{s,Q_\ell}_{kj} \le 1   \quad  \forall   k \in  {\mathcal{K}}       \label{eq:cost_cond_4}  
\end{align}  
\end{subequations}
 
All tasks must be assigned as a first or subsequent task to exactly one robot; this constraint is given by \eqref{eq:cost_cond_2}. A robot can move to, at most, one task location just after the current location, which is constrained by Eq. \eqref{eq:cost_cond_3} and  \eqref{eq:cost_cond_4}.
 
\subsection{Heterogeneous Resource Allocation Approach}
The PJITD tasks are spatio-temporal in nature. The spatio-temporal tasks need to be executed within time constraints; due to the time constraints the given number of robots may or may not be able to execute all spatio-temporal tasks. Some minimum number of robots is required in the team to execute all the given heterogeneous spatio-temporal tasks. The heterogeneous spatio-temporal tasks with given $N$ active robots may or may not be feasible due to time constraints. Hence, firstly a dynamic resource allocation algorithm \citep{velhal2022dynamic}  is modified for heterogeneous tasks (and referred as the heterogeneous resource allocation algorithm in the rest of the paper). The heterogeneous resource allocation algorithm computes the required minimum active robots with different skill sets. Once these minimum active robots from all the skill sets are available, one can guarantee that the updated heterogeneous spatio-temporal multi-task assignment problem is feasible. Then this feasible heterogeneous spatio-temporal multi-task assignment (STMTA) problem is solved by assigning the robots to multiple tasks. A robot can be set to active mode from the rest mode whenever required. Also, if a robot is not required for the execution of tasks, then it can be set to rest mode. 
 
\begin{algorithm}[t!]
 \caption{ Heterogeneous resource  allocation  algorithm } \label{algo:opti_resource}
 \begin{algorithmic}[1]
\State  {Initialize with  current $N$ robots and $M$ tasks }
\State {Solve the optimization problem (Eq. \eqref{eq:integer_prog}) } \label{algo:label:IP}
\State {$q=$ number of infeasible assignments (i.e. assignments with cost equal to $\kappa$}) 
\If {   $q>0$   }
\State Find $q$ different robots (with required skills) close to \newline  \phantom . \  \phantom . pick up locations of infeasible tasks.  i.e. \newline  \phantom . \  \phantom . $Q(R^{reserve}_i) \supseteq Q(T^{infeasible}_i )  \quad  \forall i = \{ 1,2,\cdots,q \} $
\State Add those $q$ robots in team  
\State $N = N+q $ \label{algo:label:aaa}
\State Go to step \ref{algo:label:IP}
\EndIf 
\If {All $N$ robots are not assigned to task}
\State   $p=$ number of unassigned robots  
\State   unassigned robots are set to rest mode 
\State  $ N = N-p $  
\EndIf
\State {Compute the sequence of tasks assigned to each robot using Algorithm \ref{algo:trajectory_computation} }\label{algo:label:stop}
\State{Robots that are not assigned to any task in the team are set to rest mode}
\end{algorithmic}
\end{algorithm}
  
In a heterogeneous resource allocation algorithm, at first, the algorithm solves the optimization problem (without any guarantee of a feasible solution) (step \ref{algo:label:IP} in Algorithm \ref{algo:opti_resource} ).
From the computed solution (which may or may not be feasible), the total infeasible assignments (cost equal to $\kappa$) are identified, and those many rest robots are added to active mode (steps \ref{algo:label:aaa} in  Algorithm \ref{algo:opti_resource}). While adding these reserve resources are selected such that they have the skill set to execute the considered infeasible task. The task assignment problem is solved again with the updated team of heterogeneous  robots. If the obtained solution is already feasible, then robots are assigned to tasks as per the solution. Then robots compute their respective trajectories using the trajectory computation algorithm. If any robot is unassigned, that robot is set to rest mode.
 
\subsubsection{Trajectory computations}
From the feasible solutions of the STMTA, the sequence of the tasks assigned to the robots is computed using the trajectory generation algorithm as given in Algorithm \ref{algo:trajectory_computation}. Each robot computes its own trajectory independently and follows it to execute its assigned PJITD tasks.
The trajectory generation algorithm computes the sequence of tasks assigned to the robot starting from its own position. The first assigned task defines the first two waypoints (i.e., pick-up and delivery) as given in step \ref{algo2:line6}. Note that there is no exact time constraint for the pick-up task; hence, time for pick-up is denoted with $\cdot$  (dot) to indicate feasible time (considering the time required to travel) after the previous task and before the delivery time of the current task.
 
After pick-up, the next waypoint is the spatio-temporal delivery point which defines both the delivery location and the delivery time. Once the first task is added to the trajectory, the algorithm finds if any task is assigned from the current delivery location. If a subsequent task is assigned from the delivery location, that task is augmented to the trajectory (line \ref{algo2:line9} of Algorithm \ref{algo:trajectory_computation}). This step of augmenting the subsequent task for updating the trajectory is repeated until no task is assigned from the last delivery location of the updated trajectory. 
\begin{algorithm}[htb!]
 \caption{ Trajectory Computation} \label{algo:trajectory_computation}
 \begin{algorithmic}[1]
\State {Input: task assignment solution }
\For {i = 1:N}
\State $\mu_i = \{ \} $
\If {$\sum_j \delta^{f,Q(R_i)}_{ i,j } = 1$ } 
\State $k^* = arg_k ( \delta^{f,Q(R_i)}_{i,k} = 1 )$ \label{algo2:line6}
\State $\mu_i =\left\{ ( \bm{p}^P_{k^*}, \cdot),(\bm{p}^D_{k^*}, t^D_{k^*} ) \right\}$
\While{ $\sum_j \delta^{s,Q_\ell}_{ k*,j } = 1 $} 
\State $j* = arg_j( \delta^{s,Q_\ell}_{k*,j} = 1$ ) \label{algo2:line8}
\State $\mu_i = \{\mu_i , ( \bm{p}^D_{j^*}, \cdot  ) , ( \bm{p}^D_{j^*}, t^D_{j^*} ) \}$ \label{algo2:line9}
\State $k^* = j^* $
\EndWhile
\EndIf
\EndFor
\end{algorithmic} 
\end{algorithm}
 
\section{Performance Evaluation} \label{sec:result}
The working of the proposed heterogeneous resource allocation approach for the PJITD task assignment problem is illustrated using ROS2-Gazebo simulations and lab-scale hardware experiments. 
 
\subsection{High-fidelity  Simulation Study }
\subsubsection{Simulation Setup}
The proposed use of the DREAM  algorithm for pick-up and just-in-time delivery problems is demonstrated in a simulation environment. A Gazebo simulator with \textit{RViz2}  plugins is used and operated using ROS2 (galactic) and Python. The \textit{Nav2} plugins are used for the navigation of robots in the simulations. The simulations are carried out in a Ubuntu system with an i7-8700 CPU with 16GB RAM and NVIDIA GT710 GPU.
 
\begin{figure}[t]
\centering
    \includegraphics[width=0.90\linewidth]{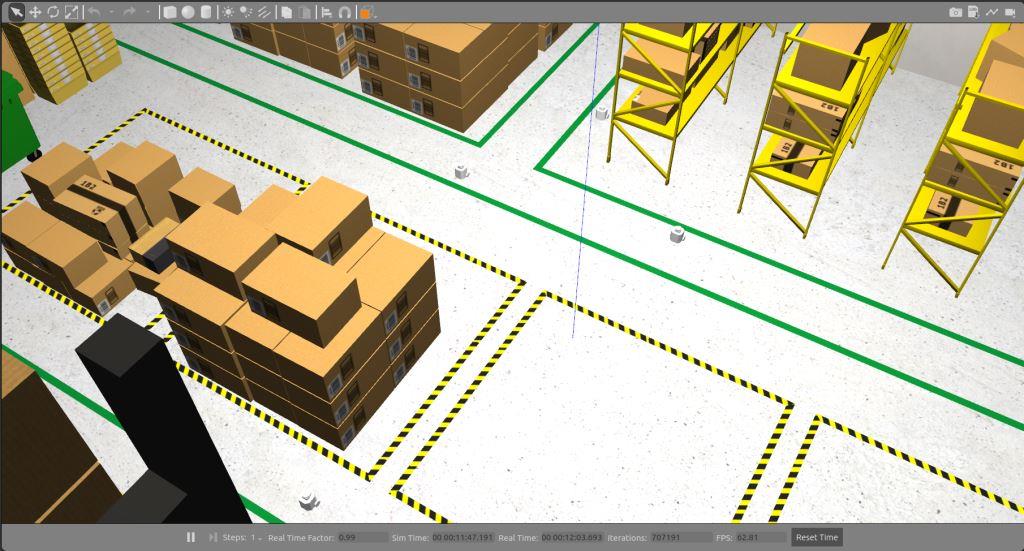}
    \caption{Warehouse model in Gazebo}
    \label{fig:warehouse} 
\end{figure}
 
The simulations are conducted for a small warehouse world designed in Gazebo;  Fig \ref{fig:warehouse} shows the warehouse. A total of four robots and seven PJITD tasks have been considered in the simulation.
 
\subsubsection{Architecture of Simulation }
\begin{figure}[b!]
    \centering
    \includegraphics[width=.95\linewidth]{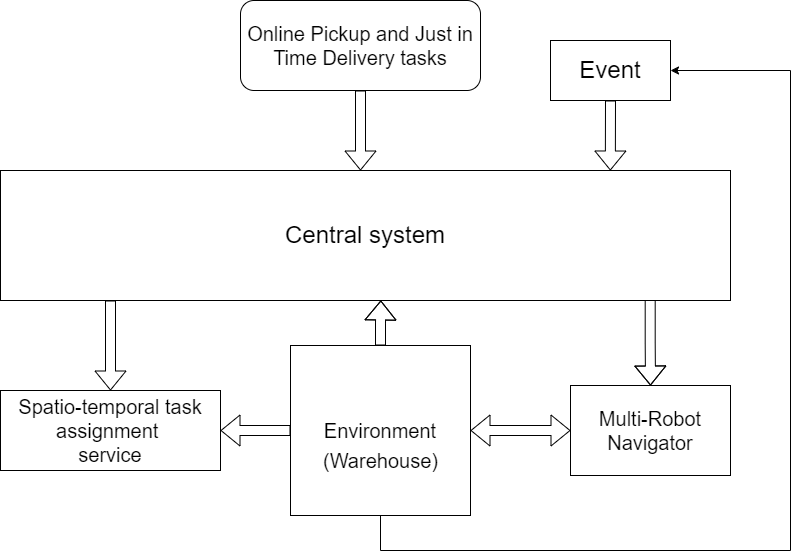}
    \caption{Architecture for simulation}
    \label{fig:architecture}
\end{figure}
Fig. \ref{fig:architecture} represents the functioning blocks of the simulation software. The architecture consists of 4 blocks: Environment, Spatio-Temporal Task Assigner, Multi-Robot Navigator, and Common Interface.
\paragraph{Environment}
The environment block corresponds to the simulation environment. It consists of the robots and all other objects in the simulation. It also provides the sensed data from each of the robots. All robots use the SLAM algorithm to localize and get the live map of the environment. Robots will operate in the environment to execute the tasks as instructed by the navigator.
\paragraph{Central system}
This software block acts as a common interface between the environment, the multi-robot navigator block, and the spatio-temporal task assignment generator block. It receives the tasks from the user/customer and sends them to the assignment generator block. Once assignments are computed, they will be received by the central system. Afterward, these assignments are shared with the navigator block to execute the tasks. Meanwhile, if any new tasks are received, the central system checks the status of the ongoing task and then updates the position of robots in future time and then calls the spatio-temporal task assignments routine. 

\paragraph{Spatio-temporal task assignment service}
All task information i.e., respective pick-up and delivery locations and delivery time, is given to the assignment service by the central system.
In the spatio-temporal task assignment service, the robots compute the navigation distances to generate the cost matrices.
The inbuilt \textit{ComputePathToPose} action-client service in \textit{Nav2} (which uses the Dijkstra's algorithm to compute the feasible path for the robot from one location to another) is used to generate the feasible trajectories. Next, the line integral along the generated trajectories provides the distances. Then cost matrices are computed  by considering the feasibility over desired time using Eq. ~\eqref{eq:cost_cf} and ~\eqref{eq:cost_cs}. Then the optimization problem defined by Eq. ~\eqref{eq:integer_prog} is solved using the \textit{linear\_sum\_assignment} function from optimize library in scipy. Next, each robot's trajectories are computed for obtained assignments using the trajectory generation algorithm. These trajectories are returned as the output to the central system.
\paragraph{Multi-robot navigator}
Once the central system receives the trajectories for each robot, it is sent to the navigation service. 
The \textit{Nav2} plugin is used in the navigation node to navigate each robot. For simultaneous operations of multiple robots, \textit{Nav2} services are called asynchronously. Each robot will travel along its assigned trajectory. For the individual robot, a task is subdivided into four sub-tasks: reaching to pick-up station, loading items, reaching the delivery station, and unloading the items at desired delivery time. All these sub-tasks are executed sequentially (synchronously). However, in a team sense, all robots operate asynchronously to execute their individual tasks simultaneously.

\paragraph{Event}
Suppose something abrupt happens in the environment, such as the failure of one robot or the closure of some roadways. Then the event is triggered. After the event, the central system checks the status of all active tasks and solves the task assignment problem again for scenarios after the event. Navigation services are updated, and robots are assigned to tasks according to the updated assignments

\subsubsection{Simulation Results}
\begin{table}[]
 \centering
 \setlength{\arrayrulewidth}{1pt}
\begin{adjustbox}{max width=0.95\linewidth} 
\renewcommand{\arraystretch}{1.0}
\begin{tabular}{|c|c|c|c|c|c|c|c|}
\hline
\multicolumn{1}{|l|}{}  & $T_0$  & $T_1$     & $T_2$   & $T_3$    & $T_4$   & $T_5$  & $T_6$ \\ \hline
\begin{tabular}[c]{@{}c@{}}Pick up \\ location\end{tabular}  & 4.0,1.0 & 1.0,-3.5 & -4.0,0.5 & 1.5,5.0  & 1.0,3.5  & 2.0,1.0 & 2.0,2.0 \\ \hline
\begin{tabular}[c]{@{}c@{}}Delivery \\ location\end{tabular}  & 1.0,2.5  &-2.5,-2.5 & 1.0,-2.0   & -4.0,2.5 & 4.0,-2.5 & 0.0, 2.5 & 4.0, 2.0  \\ \hline
\begin{tabular}[c]{@{}c@{}}Delivery \\ time\end{tabular}  & 35 & 50 & 60 & 75 & 100 & 120 & 140      \\ \hline
\begin{tabular}[c]{@{}c@{}} loading  \\ time ($\tau^l$) \end{tabular} &1 & 1 & 1& 2& 2& 1& 1   \\ \hline
\begin{tabular}[c]{@{}c@{}} unloading  \\ time ($\tau^u$) \end{tabular} & 1 & 2 & 1& 1& 2& 1& 1   \\ \hline
\end{tabular}
\end{adjustbox}
 \vspace{0pt}
 \caption{The pick-up and delivery locations with respective  delivery, loading, and unloading times for the considered tasks}
 \label{table:simulation_tasks}
 \end{table}
 
The simulations are conducted with seven tasks ($T_0$ to $T_6$) and four robots ($R_0$ to $R_3$)  serving those tasks to illustrate the working of the proposed heterogeneous resource allocation approach for PJITD tasks. The PJITD tasks are given in table \ref{table:simulation_tasks}. The robots are initialized at $R_0(0) =(2,-0.35) $, $R_1(0) = (1.6,2.5)$, $R_2(0) = (-3.0,1.2)$, and $R_3(0) = (3.6,1.5) $.
To visualize the coordinates of the pick-up and delivery points, unique color is allocated to each task, where a circle shows the pick-up point, and the delivery point is shown by a square dot along with its delivery time. The current simulation time is displayed at the top right corner. 
The simulation video is available at \url{https://www.youtube.com/watch?v=gNC0hG4CG2A}.

\begin{figure}[tbp!]
    \centering
    \includegraphics[width=0.9\linewidth]{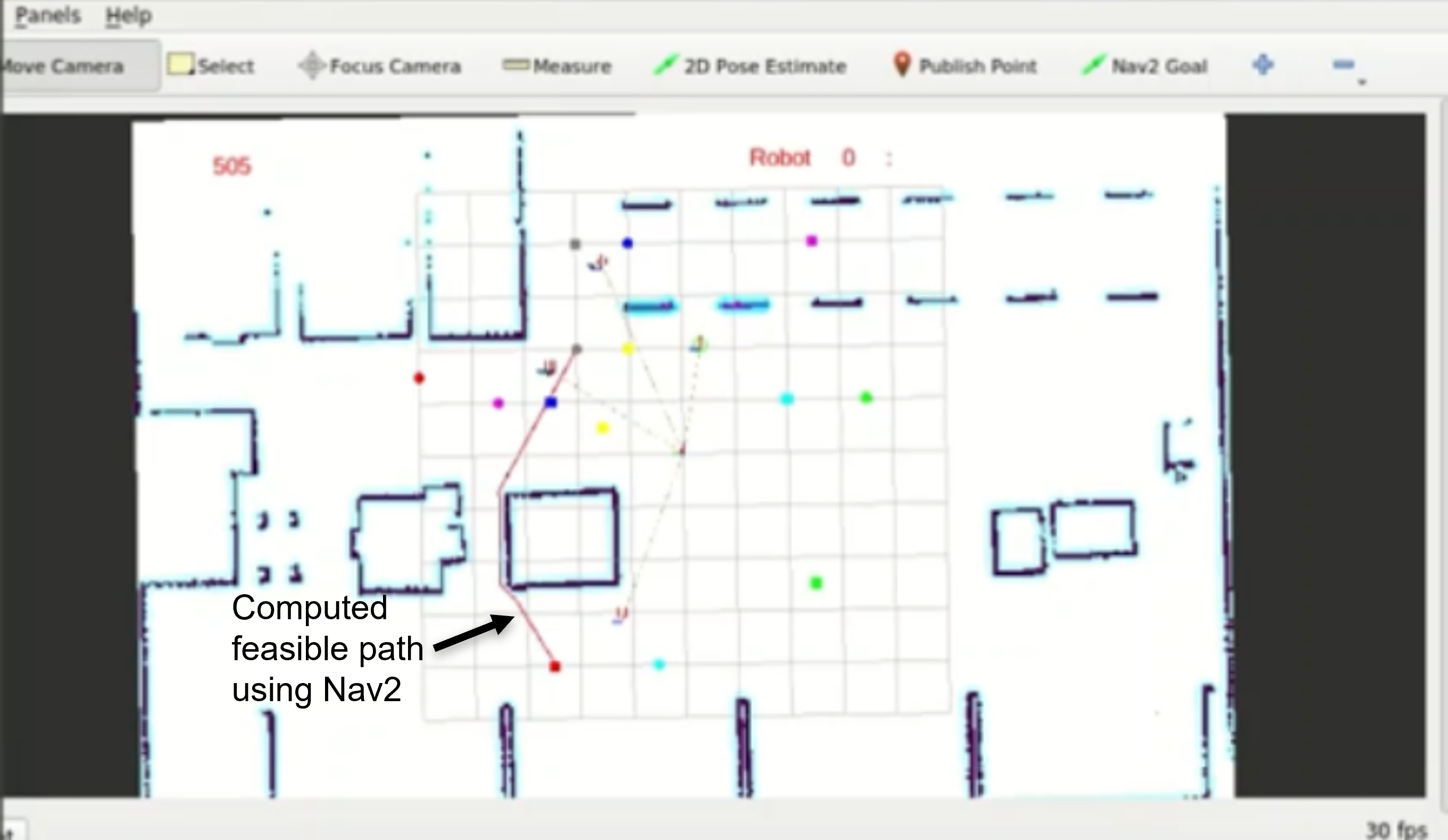}
    \caption{Typical of the feasible path computed using Nav2}
    \label{fig:compute_path}
\end{figure}
Each RViz2 window is dedicated to each robot to show its operations and status. The action of each robot is decided using the proposed algorithm and displayed at the top of the corresponding RViz2 window. The local navigation trajectory computed by the robot is shown in a red colored curve. Once the navigation is initiated, the status indicates which task the robot is executing. It indicates whether the robot is executing the pick-up, delivery, loading, or unloading subtask. The status of a robot is updated after each operation. Once all the tasks allocated to a specific robot are completed, the same is highlighted in the status.

The robots are assigned tasks for the given PJITD tasks using the proposed method. For that purpose,  robots compute the path distances for the given tasks. \textit{ComputePathToPose} function from the \textit{Nav2} package has been used to compute the feasible path from one point to another point in the arena. The RViz2 snip-shot of the warehouse is shown in Fig \ref{fig:compute_path}, where it also shows the feasible path between two points by a red-colored curve. The line integral along this feasible path is used to compute the distances. 

\begin{figure*}[t]
\centering
    \begin{subfigure}[b]{0.45\linewidth}
        \includegraphics[width=1\linewidth]{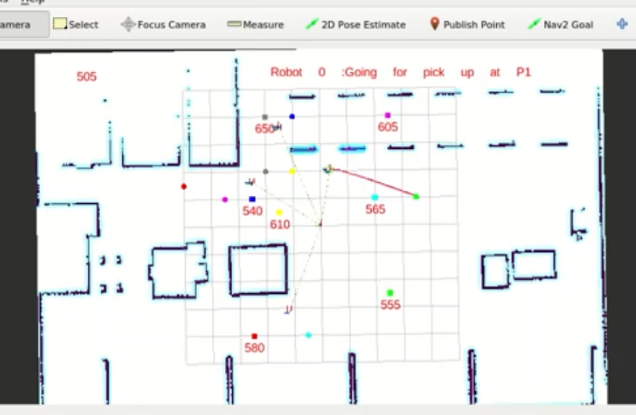} 
        \caption{Robot 0}  
        \label{fig:subfig_sim1}
    \end{subfigure} %
    \quad
    \begin{subfigure}[b]{0.45\linewidth}
        \includegraphics[width=1\linewidth]{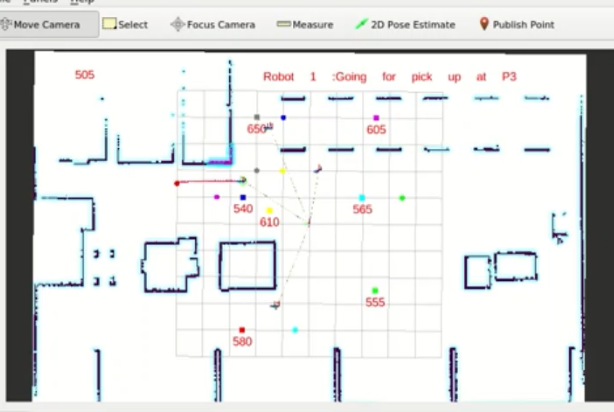} 
        \caption{Robot 1}   
        \label{fig:subfig_sim2}
        \end{subfigure}\\
    \vspace{10pt}
    \begin{subfigure}[b]{0.45\linewidth}
        \includegraphics[width=1\linewidth]{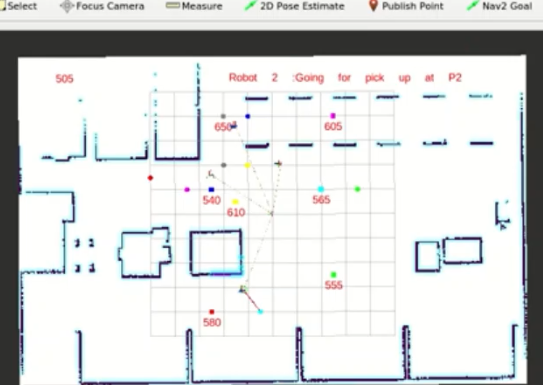} 
        \caption{Robot 2} 
        \label{fig:subfig_sim3}
    \end{subfigure} %
    \quad
    \begin{subfigure}[b]{0.45\linewidth}
        \includegraphics[width=1\linewidth]{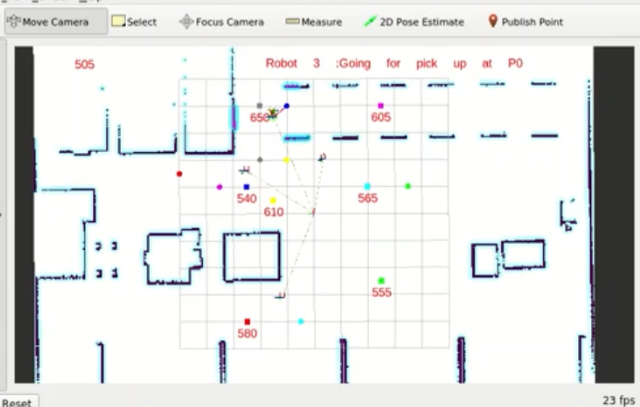} 
        \caption{Robot 3}  
        \label{fig:subfig_sim4}
    \end{subfigure}\\
\caption{Initial assignments for four robots.  \textnormal{ Each sub-figure shows the robots moving from their initial location to their first assigned pick-up locations at the start of simulations at t = 505 sec.} }
\label{fig:sim_t_505}
\end{figure*}

The cost matrix is computed for given tasks and robot positions using Eq. ~\eqref{eq:cost_cf}, and ~\eqref{eq:cost_cs} are given below by Eq. \eqref{eq:c_f_gazebo} and \eqref{eq:c_s_gazebo} respectively.
The robots considered in simulations are homogeneous and have all skill sets to execute all tasks; so, the subsequent cost matrix is the same for all four robots, i.e.  $ C^{s,0} = C^{s,1} =C^{s,2} =C^{s,3} = C^{s,\bullet}$. 
The solution obtained from the proposed heterogeneous resource allocation approach is the decision variable indicating the robot positions assigned to the tasks. The box in cost matrices highlights the task assignment solution, and the superscript denotes the robots to which the task is assigned. 

\begin{align}
 \begin{bmatrix*}[r]  {\bf{C}}^{f,0} \\ {\bf{C}}^{f,1} \\ {\bf{C}}^{f,2} \\ {\bf{C}}^{f,3}   \end{bmatrix*}= & \resizebox{.75\linewidth}{!}{$ \begin{bmatrix*}[c]   
 5.66   & \fbox{6.95}^0    & 11.73  & 11.44  & 10.61  & 2.84   & 4.25   \\
6.13   & 9.67   & 22.72  &  \fbox{8.56}^1   & 7.77   & 3.11   & 2.60   \\
10.54  & 9.78   &  \fbox{6.75}^2    & 12.77  & 12.01  & 6.69   & 7.21   \\
\fbox{3.94}^3   & 9.28   & 13.41  & 10.30  & 9.88   &  \  3.25   &  \    3.58    
\end{bmatrix*} $}  \label{eq:c_f_gazebo} \\
 {\bf{C}}^{s,\bullet} = &  \resizebox{.750\linewidth}{!}{$ \begin{bmatrix*}[c]  
\ \      \infty  \ \  &  \ \  9.62   & \ \  11.12  & \quad 8.60   & 7.62   & 3.36   & 3.08   \\
\infty    &       \infty & 8.91   & 14.56  & 13.56  & 7.17   & 8.24   \\
\infty & \infty &      \infty & 1.31   & 12.12  & \fbox{4.65}^2    & 6.02   \\
\infty & \infty & \infty &      \infty &\fbox{11.96}^3   & 8.15   & 8.67   \\
\infty & \infty & \infty & \infty &    \infty & 100    & 6.87   \\
\infty &  \infty &   \infty &   \infty &   \infty & \    \infty &    \fbox{3.49}^2 
\end{bmatrix*} $ }  \label{eq:c_s_gazebo}
\end{align}

Next, the trajectories of each robot are obtained using the trajectory computation algorithm (Algorithm \ref{algo:trajectory_computation}). The obtained trajectories for $R_0$ is $\mu_0 = \{T_1\}$ ,for $R_1$ is $\mu_1 = \{T_3\}$, for $R_2$ is $\mu_2 = \{T_2,T_5,T_6\}$, and for $R_3$ is $\mu_3 = \{T_0, T_4\}$.

The task execution starts at the simulation time of $505 \ sec$, accordingly, all the delivery times are updated. Fig \ref{fig:sim_t_505} shows the initial scenario at $t = 505 sec$, where all four robots are assigned to their respective tasks. It also shows the robots' operations in their respective RViz2 windows.  

Each robot has to execute the assigned tasks consisting of reaching the pick-up location, loading items, going to the delivery location, waiting till the desired delivery time, and then unloading the items. Each robot executes these sub-tasks sequentially and then executes the subsequently assigned tasks.
Robot $R_0$ has been assigned to task $T_1$. $R_0$ moves towards the pick-up location $P_1$ (denoted by a green dot). At the same moment, $R_1$ is assigned to task $T_3$, so it moves for pickup at $P_3$. Similarly, $R_2$ moves towards the pickup at $P_2$, and $R_3$ moves towards the pickup at $P_0$.
Fig \ref{fig:sim_t_505}, shows the snapshot at t= 505 (tasks are given to the algorithm at 505 sec.), where robots find the feasible trajectories to reach respective pick-up locations. 

Robot $R_3$ reaches its pickup location $P_0$ at $t=513$ sec, it waits there for $1$ sec to load the items. Next $R_3$ moves towards its delivery location $D_0$ and reaches $D_0$ at $t= 531$ sec. The delivery time of task $T_0$ is $540$, so it waits for the next 9 seconds. Meanwhile, other robots $R_0$,$R_1$, and $R_2$ had reached respective pick-up locations, loaded the items, and are traveling towards their respective delivery locations.
At $t = 540$ sec, $R_3$ unloads the items in $1$ sec. So the $R_3$ completes the task $T_0$ at $541$ sec. After completing first assigned task, $R_2$ starts executing its next task $T_4$. For that purpose, it travels towards $P_4$ for pick-up.

The $R_0$ reaches the  delivery location $D_1$ at  $546$ sec and the delivery time for $T_1$ is $555$ sec; so it waits for 9 seconds. After that, it unloads the items in 1 sec. As $R_0$ is assigned to only one task, it goes to rest mode after completing it.
The $R_2$ reaches its first delivery location $D_2$ at $t = 551$, it waits till the delivery time, i.e.  $565$ sec, and unloads the item in the next $1$ sec. So, $R_2$ competes the task $T_2$ at $t = 566$. After completing the first assigned task $T_2$, $R_2$ starts its next task i.e. task $T_5$.
Meanwhile,  $R_1$ reaches its first delivery location $D_3$ at $t = 556$,  so it waits there till $t = 580$, and then unloads the items in $1$ sec. As $R_1$ is assigned to only one task, it goes to rest mode after completing all tasks.

Now only two robots, $R_2$ and $R_3$, are active and executing their tasks. $R_0$ and $R_1$ are in rest mode. $R_3$ reaches its assigned delivery $D_4$ at t = 591 and waits until the given delivery time of $605$ sec.  After t = 605, $R_3$ unloads the items in $2$ seconds and completes its last task $T_4$, and goes to rest mode. $R_2$ reaches to the delivery location $D_5$ at $t = 599$, waits till its delivery time, and after $t = 610$ unloads the items and starts its new task $T_6$. $R_2$ picks up the item from $P_6$ and reaches $D_6$ at $t = 645$. After waiting for $5$ sec, it unloads the item and completes its last task.
From the simulation video, one can observe that robots are able to execute all the given PJITD tasks. 

\subsubsection{Computational Complexity} 
Table \ref{table:comp_time} shows the computational time required for the proposed approach for PJITD task assignments. The cost of the tasks is computed using the map computed for the given warehouse; the cost matrix computation takes significant time. The algorithm is computationally efficient and requires time almost three orders less than the cost matrix computation. 
From the table, one can observe that as the number of tasks increases, the total computational time increases, and the total computational time is quadratically proportional to the number of tasks.

 \begin{table*}[htbp!]
  \centering
\begin{adjustbox}{width=1\textwidth}
\begin{tabular}{|l|r|r|r|r|r|r|r|r|r|r|r|r|}
\hline
\begin{tabular}[c]{@{}l@{}} Number \\ of tasks\end{tabular}                      & 10     & 20     & 30     & 40     & 50     & 60     & 70     & 80     & 90     & 100   & 200 & 500                      \\ \hline 
\begin{tabular}[c]{@{}l@{}} Average \\ computation \\ time  (msec)\end{tabular} & 0.0481 & 0.1283 & 0.2837 & 0.5294 & 0.8929 & 1.3869 & 2.0116 & 2.7918 & 3.7553 &  4.9230 &    31.5588 & 408.9476\\ \hline 
\end{tabular} 
\end{adjustbox}
 \caption{Average computational time for solving optimization problem (Eq\eqref{eq:integer_prog})for different number of tasks }
 \label{table:comp_time}
\end{table*}

\subsection{Resource Utilization } 
The proposed approach requires a dynamic number of robots to execute the given spatio-temporal tasks. For this purpose, the resources (robots) required to execute given spatio-temporal tasks for the different arrival rates of the tasks are analyzed.  Here, PJITD tasks are generated at random pick-up and delivery locations with delivery times generated with the Poisson distribution.
The resource utilization factor (RUF)  for a given team size ($n$) is defined as the ratio of the total time interval at which a team of $n$ robots is active to the total simulation time and is defined as
\begin{align} 
\text{RUF}(n) = \frac{\text{time interval with n robots }} {\text{Total simulation time}} \times 100
\end{align}

\begin{figure}[t]
    \centering
    \includegraphics[width=\linewidth]{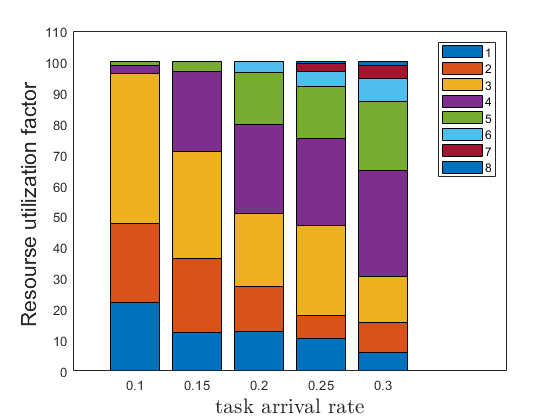}
    \caption{Resource utilization with different arrival rates of PJITD tasks }
    \label{fig:RUF}
\end{figure}

Fig \ref{fig:RUF} shows the RUF for the different arrival rates ($Q$) of spatio-temporal tasks. 
For an arrival rate of 0.1, at max 5 robots are required and out of which, five robots are required only for the $1.15\% $, four robots are required for $2.68\% $, three robots are required for $48.35\%$, two robots are required for $25.61\%$ and only one robot is required for $22.19\%$. 
From Fig \ref{fig:RUF}, one can observe that as the arrival rate increases the number of robots required increases.

\subsection{Implementability Study } 
\begin{figure}[t]
    \centering
    \includegraphics[width=\linewidth]{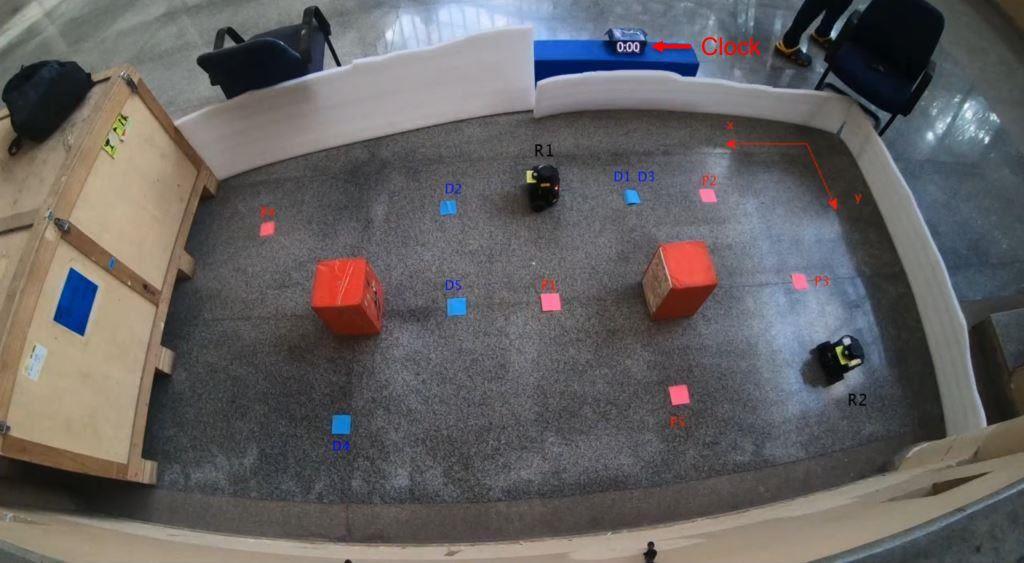}
    \caption{Arena for hardware experiment }
    \label{fig:arena}
\end{figure}

The proposed approach in this paper computes the required number of heterogeneous robots and their assignments to execute given tasks. To illustrate the implementability of the proposed non-iterative and online computable solution for heterogeneous PJITD tasks is demonstrated in the lab scale experiment. Fig \ref{fig:arena} shows the arena considered in the experiments. The arena is of dimensions $ 3.6 m \times 2 m $ with the origin on the right top. Two cuboid shaped obstacles are added to the arena at $ \bm{O}_1 = ( 1.37, 0.96) $ and $ \bm{O}_2 =(2.78,0.96) $. 
Two TurtleBot3 Burger robots equipped with Raspberry Pi 3B+ and Ubuntu 18.04 Server with ROS Melodic  \footnote[1]{hardware support for ROS2 is still in development, so we have used ROS melodic framework for hardware implementation } have been used for hardware experiments. 
The map of the arena is generated using the Simultaneous Localisation and Mapping (SLAM) algorithm on the  LDS-01 LIDAR sensor outputs obtained by teleoperation of the Turtlebot in the arena.
For the turtlebot's navigation, an inbuilt  \textit{Nav} package is used by each turtlebot.
\begin{figure}[]
    \centering
    \includegraphics[width=\linewidth]{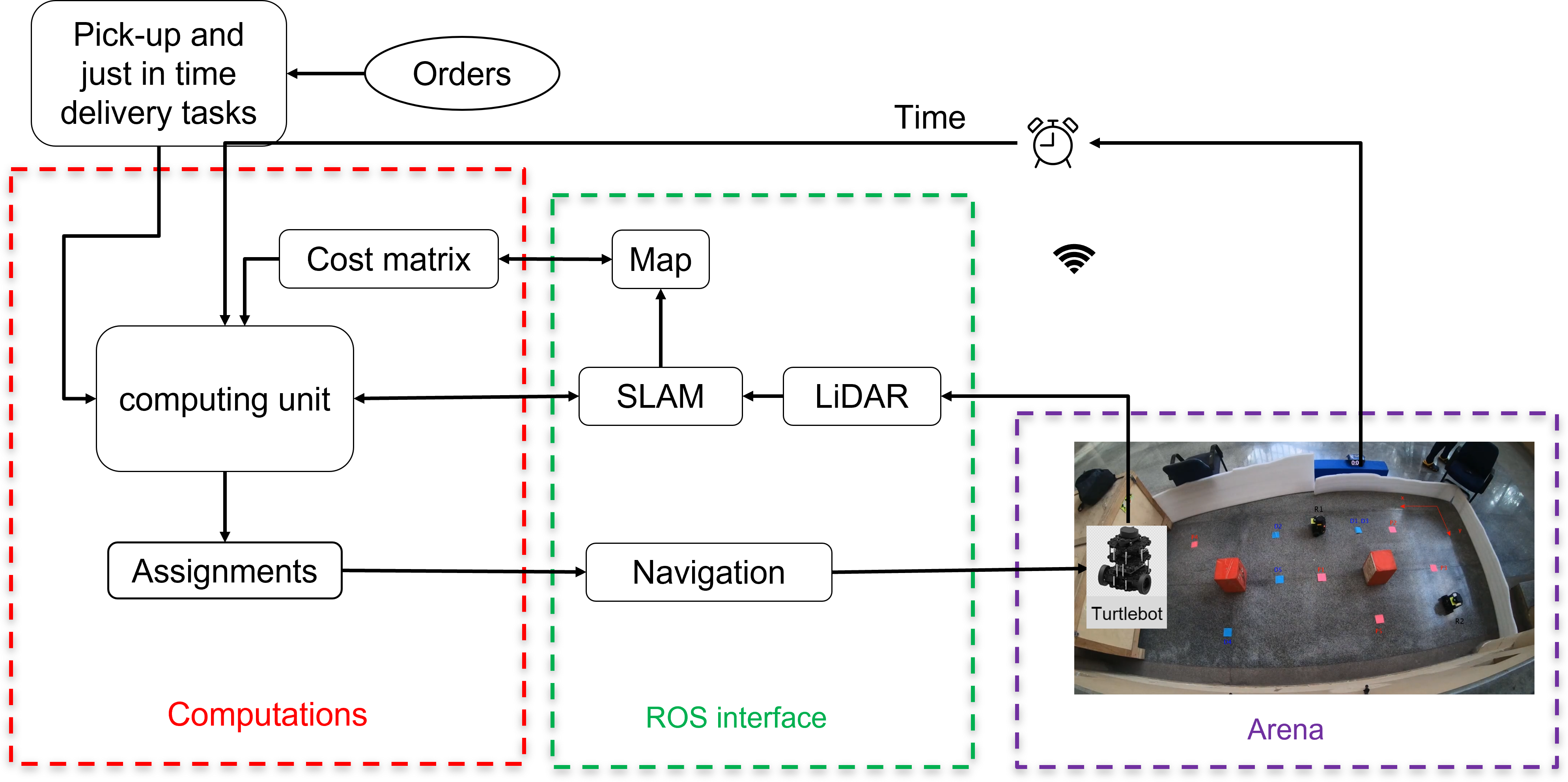}
    \caption{Architecture of hardware experiment }
    \label{fig:architecture_hardware}
\end{figure}

At the start of experiments, the turtlebots are placed in the arena, and PJITD tasks are defined. Using the map and task information centralized algorithm computes the sequence of tasks assigned to each turtlebot, and the sequence of tasks is communicated to each turtlebot. Navigation ROS Node is used to navigate the robot to its desired waypoints in the proper order while keeping the temporal constraints satisfied.

Fig. \ref{fig:arena} shows the arena with given pick-up and delivery points. Additionally, a real-time clock is placed near the arena's top right corner to highlight the experiments' temporal features. The pick-up delivery tasks for the experiments are given in table \ref{table:experiment_tasks}. The robots are initialized at (1.7,0.5) and (0.4,1.12) for robot 1 and robot 2, respectively. The Turtlebot3 can have a maximum speed of $0.22m/sec$. However, practically it has been observed that for a small arena of this size, the practical speed reached is $0.12m/sec$ due to acceleration and deceleration limits. For this experiment, the robot's velocity is taken as  $0.12m/sec$.

\begin{table}[t]
 \centering
 \setlength{\arrayrulewidth}{1pt}
\begin{adjustbox}{max width=0.95\linewidth} 
\renewcommand{\arraystretch}{1.1}
\begin{tabular}{|c|c|c|c|c|c|}
\hline
\multicolumn{1}{|l|}{}    & $T_1$     & $T_2$   & $T_3$    & $T_4$   & $T_5$  \\ \hline
\begin{tabular}[c]{@{}c@{}}Pick up \\ location\end{tabular}  & (1.9,0.85 ) & (0.8,0.35)  & (0.5,0.85) & (3.0,0.4)  & (1.2,1.3 ) \\ \hline
\begin{tabular}[c]{@{}c@{}}Delivery \\ location\end{tabular} & (1.2, 0.35) & (2.0,0.35 ) & (1.2,0.35) & (2.5,1.25) & (2.0,0.85) \\ \hline
\begin{tabular}[c]{@{}c@{}}Delivery \\ time\end{tabular}     & 35          & 50          & 90         & 110         & 120        \\ \hline
\end{tabular}
\end{adjustbox}
 \vspace{0pt}
 \caption{The pick-up and just-in-time delivery tasks}
 \label{table:experiment_tasks}
 \end{table}

Now, the robot computes the path distances for the given tasks. \textit{Nav} package has been used to compute the feasible path from one point to another point in the arena. The line integral along this feasible path is used to compute the distances. The cost matrix computed is computed using Eq. ~\eqref{eq:cost_cf} and ~\eqref{eq:cost_cs} and given as,
\begin{align}
\begin{bmatrix*}[r] {\bf{C}}^{f,1} \\ {\bf{C}}^{f,2} \end{bmatrix*} &= \resizebox{.750\linewidth}{!}{$ \begin{bmatrix*}[c]
 \fbox{1.7278}^1 & 3.0593   & 2.5471 & 4.3748  & 2.3141 \\
2.4100  & \fbox{3.0920}^2   & 1.4627   & 5.9802 & 2.1106  \\     \end{bmatrix*}  $} \label{eq:c_f}     
\end{align}

\begin{align}
{\bf{C}}^{s,1} &=  \resizebox{.750\linewidth}{!}{$\begin{bmatrix*}[c]
\quad  \infty \quad \  & 1000 &   \fbox{1.9985}^1       & 4.9158       & 2.6336  \\
\infty & \quad \infty \quad \  &  2.8275   & 4.0760     & 2.6032   \\
\infty & \infty & \infty & 1000    & \fbox{2.6336}^1  \\
\infty & \infty & \infty & \infty & 1000   \\
\infty & \infty & \infty & \infty & \infty     \end{bmatrix*} $}   \label{eq:c_s,1}  \\[3pt] 
{\bf{C}}^{s,2} &=  \resizebox{.750\linewidth}{!}{$ \begin{bmatrix*}[c]
\quad \infty \quad \  & 1000 &   1.9985       & 4.9158       & 2.6336  \\
\infty &  \infty   &  2.8275   & \fbox{4.0760}^2      & 2.6032   \\
\infty & \infty & \infty & 1000    & 2.6336 \\
\infty & \infty & \infty & \infty & 1000   \\
\infty & \infty & \infty & \infty & \infty     \end{bmatrix*} $} \label{eq:c_s,2}
\end{align}

\begin{figure*}[]
\centering
\begin{subfigure}[b]{0.48\linewidth}
\includegraphics[width=1\linewidth]{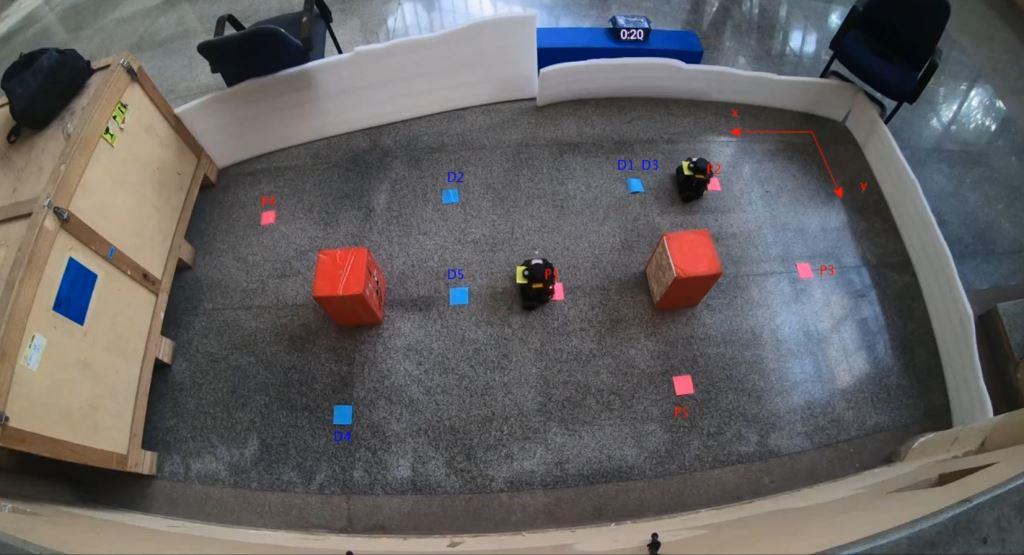} 
\caption{t = 20 sec}  
\label{fig:subfig1}
\end{subfigure} %
\begin{subfigure}[b]{0.48\linewidth}
\includegraphics[width=1\linewidth]{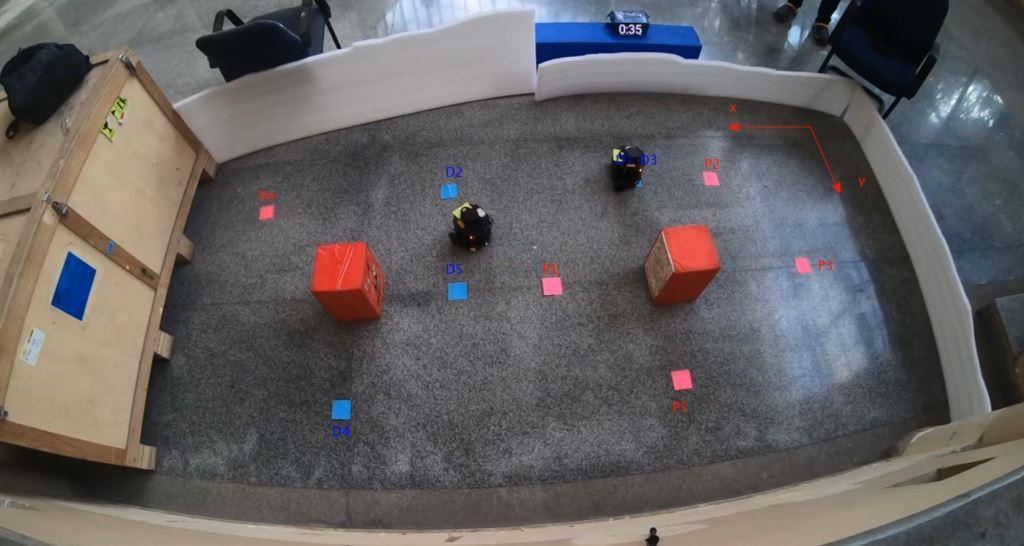}
\caption{t = 35 sec}   
\label{fig:subfig2}
\end{subfigure}\\
\begin{subfigure}[b]{0.48\linewidth}
\includegraphics[width=1\linewidth]{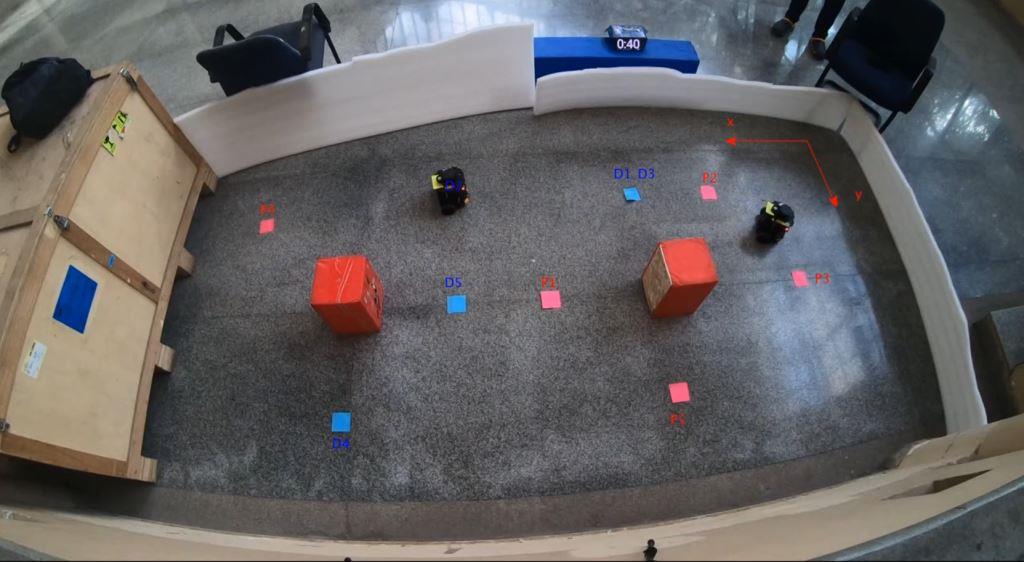} 
\caption{t = 40 sec} 
\label{fig:subfig3}
\end{subfigure} %
\begin{subfigure}[b]{0.48\linewidth}
\includegraphics[width=1\linewidth]{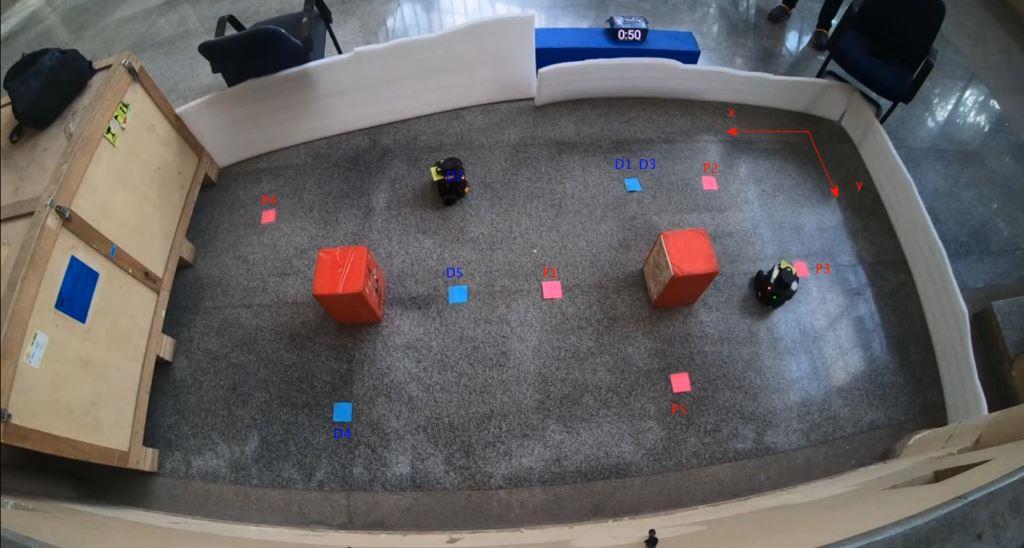}
\caption{t = 50 sec}  
\label{fig:subfig4}
\end{subfigure}\\
\vspace{5pt} 
\begin{subfigure}[b]{0.48\linewidth}
\includegraphics[width=1\linewidth]{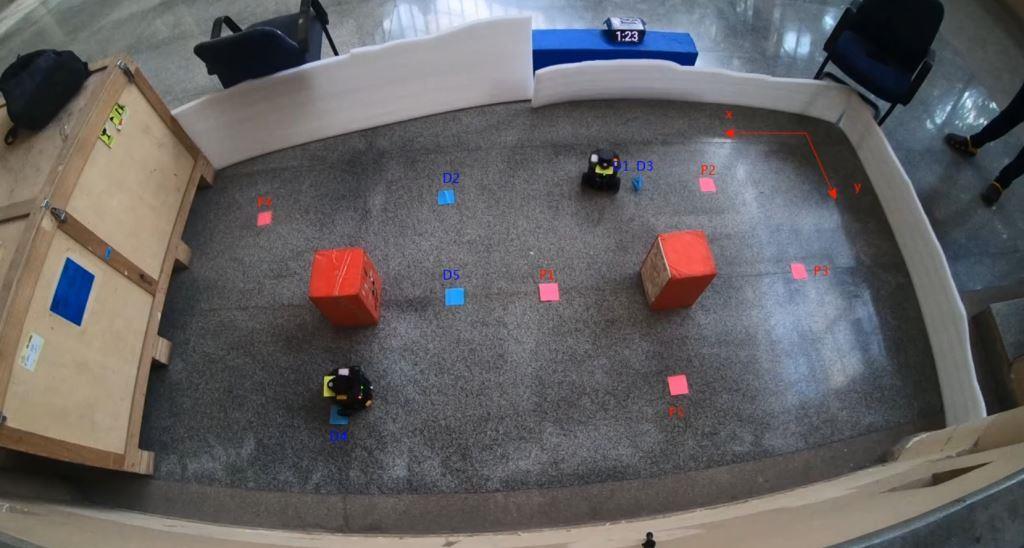} 
\caption{t = 83 sec}  
\label{fig:subfig5}
\end{subfigure} %
\begin{subfigure}[b]{0.48\linewidth}
\includegraphics[width=1\linewidth]{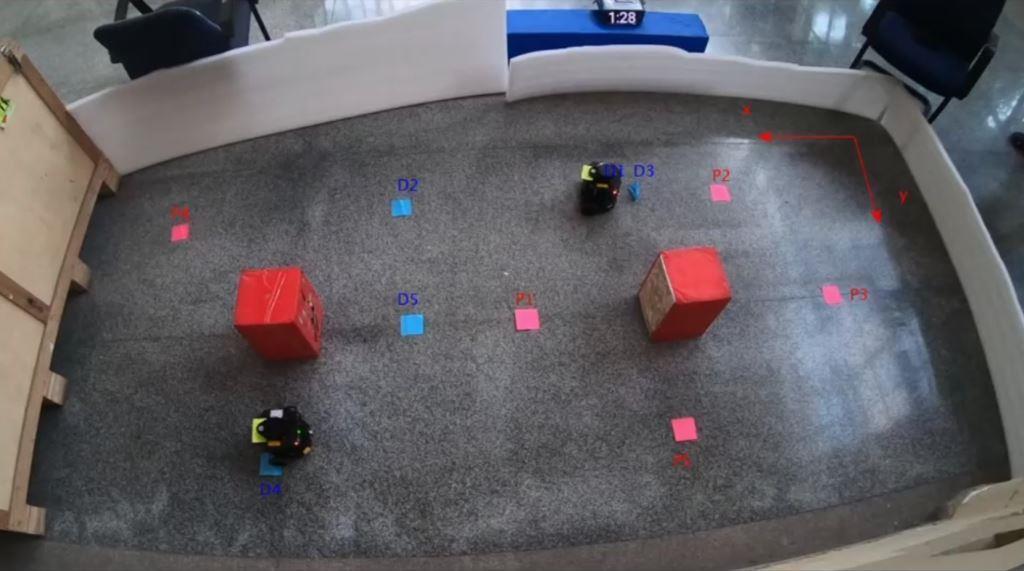}
\caption{t = 88 sec}  
\label{fig:subfig6}
\end{subfigure}\\
\begin{subfigure}[b]{0.48\linewidth}
\includegraphics[width=1\linewidth]{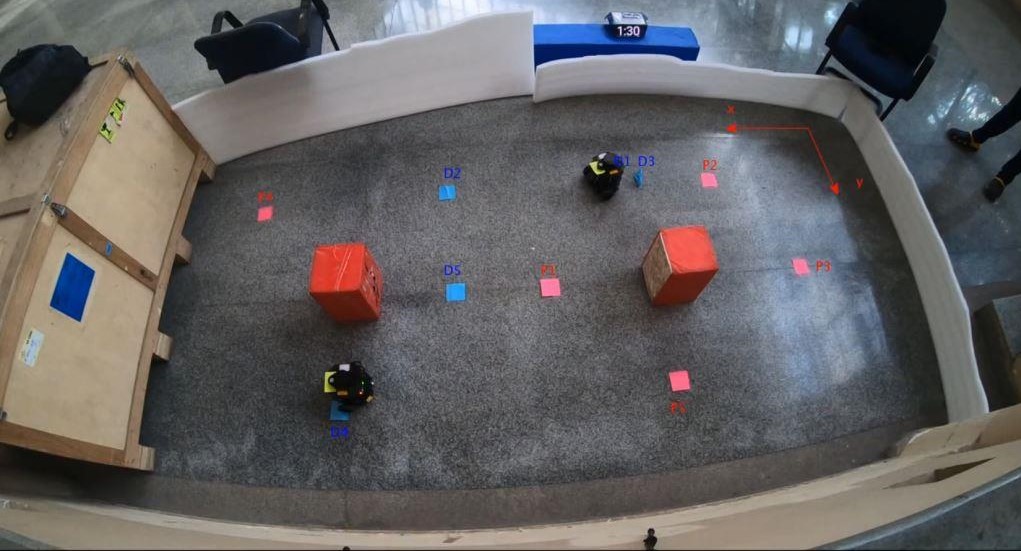} 
\caption{t = 90 sec}  
\label{fig:subfig7}
\end{subfigure} %
\begin{subfigure}[b]{0.48\linewidth}
\includegraphics[width=1\linewidth]{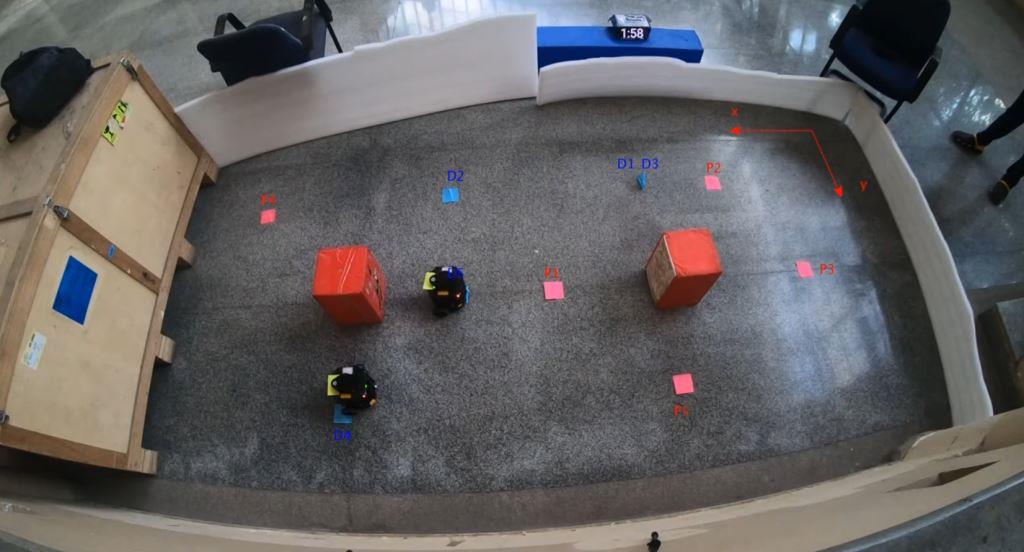} 
\caption{t = 118 sec}  
\label{fig:subfig_intruder}
\end{subfigure}\\
\vspace{5pt} 
\caption{Snapshots of the turtlebots executing the PJITD tasks at various time instances. \textnormal{ 
a) $t= 20$, $R_1$ is at pick-up $P_1$ and $R_2$ is at pick-up  $P_2$. 
b) $t = 35$ $R_1$ completes its delivery at $D_1$. 
c)  $t = 40 $ $R_2$ reaches at $D_2$ and waits till delivery time of 50 sec.
d) $t = 50$ $R_2$ completes its delivery at $D_2$. 
e) t = 83, $R_2$ reaches to $D_4$ and  waits. 
f) t = 88, $R_1$ reaches to $D_3$ and  waits 
g) t = 90, $R_1$ delivers at $D_3$. 
h) t = 118, $R_1$ reaches to $D_5$.}}
\label{fig:hardware_expt_result}
\end{figure*}

Note that, the diagonal and lower triangular elements of the cost matrix $C^{s,1}$ and $C^{s,2}$   are set to $\infty$ as these tasks happened in the past. The forward-time infeasible assignments are given a large value of 1000. For the element $C^{s,1}(1,2)$ i.e. the robot $R_1$ executing the task $T_2$ after the execution of task $T_1$ the total time available is ($t^D_2 - t^D_1 = 15$) and the distance needs to be traveled is $0.48$ for pickup and $2.05786$ for delivery. The total distance that needs to be traveled is $2.5378$, and it requires a minimum time of $21.48 sec$. But the available time ($15 sec$) is less than the minimum time required; hence the cost is set to 1000. Similarly the cost $C^{s,1}(3,4)$ and $C^{s,2}(4,5)$ are also set to $1000$.

The solution obtained from the DREAM approach is the decision variable indicating the robot positions assigned to the tasks. In cost matrices, the task assignment solution is marked by boxes.
Next, the trajectory generation algorithm is used to compute the trajectory of each robot. The obtained trajectories for turtlebot 1 and turtlebot 2 are $\mu_1 = \{T_1,T_3,T_5\}$ and $\mu_2 = \{T_2,T_4\}$ respectively.

The hardware experimental run is video-graphed and the video is available  at \url{https://www.youtube.com/watch?v=wuwL5-OVjyM}.  Here, we explain the experiment with Fig \ref{fig:hardware_expt_result}.
$R_1$ has  been assigned to the tasks $T_1, T_3, T_5$ and R2 has been assigned to the tasks $T_2,T_4$. Robots need to execute these tasks in sequence. As per the assignments, at $t = 0$, $R_1$ starts executing the $T_1$ as per assignment and $R_2$ starts the task $T_2$.  $R_1$ navigates to $P_1$ and $R_2$ navigates to $P_2$ for pick-up. After reaching the pick-up location, the turtlebot aligns its heading towards the positive x direction (towards the left in the video). After pick-up, the turtlebot moves toward the respective delivery location. The turtlebot 1 reaches the delivery location $D_1$ and aligns to the positive x direction at $t=35  sec$. The desired delivery time at $D_1$ is also $35 sec$; the turtlebot has reached on time, and task $T_1$ is executed successfully. 

Next the turtlebot $R_1$ starts executing its next task  $T_3$ and moves towards the pick-up location $P_3$. Meanwhile, $R_2$ reaches it delivery location $D_2$ at $ t= 40$; but desired delivery time of $T_2$ is $50 sec$, so $R_2$ waits at $D_2$ for $10$ seconds and at $t= 50$ $R_2$ completes the task $T_2$. Then $R_2$ starts its next task $T_4$ by moving towards pick-up location $P_4$. $R_2$ reaches its delivery location $D_4$ at $t= 83$ sec. The delivery time of $T_4$ is 110 sec, hence it waits at $D_4$ for the next $27$ seconds and completes the task $T_4$.

Meanwhile, $R_1$ which is executing task $T_3$ reaches the delivery location $D_3$ at $t = 88 sec$, waits for $2$ sec till its desired delivery time, and completes the task $T_3$ at $t = 90 sec$. After completing task $T_3$, $R_1$ starts its next and last task $T_5$, it reaches to $P_5$ for pick-up and then moves toward the delivery location $D_5$. $R_1$ reaches the $D_5$ at $t=119 sec$. $R_1$ waits for 1 sec and completes its last task at the desired delivery time of $120 sec$. One can observe that the feasible tasks were assigned to turtlebots by the proposed algorithm, and turtlebot executes the tasks on time.

\section{Conclusion} \label{sec:conclusion}
In this paper, we propose a non-iterative heterogeneous resource allocation for multi-task assignments to handle the online Pickup and Just-In-Time Delivery (PJITD) tasks with heterogeneous robots.
In PJITD tasks, delivery time is constrained to the desired time, so the PJITD problem is formulated as the heterogeneous spatio-temporal multi-task assignment (STMTA) problem. The cost function of STMTA has been modified to include the traveling time, operating time, and heterogeneous skills required for the task.
The proposed heterogeneous resource allocation approach utilizes the robots minimally to execute all the given heterogeneous PJITD tasks, and obtained assignments are optimal (minimum the total distances traveled by the team of robots). The working of the approach has been demonstrated using high-fidelity simulations and hardware implementation. The PJITD tasks can be assigned to robots/agents by online computation, and the same has been demonstrated using high-fidelity simulations and hardware experiments.
Future work will explore the use of spatio-temporal task assignment formulation for various applications which were discarded due to the unavailability of online solutions. Also, the proposed approach will be studied for combining the scheduling and spatio-temporal task assignment problems in future works.  

\section*{Acknowledgment}
The authors would like to thank Nokia Centre for Excellence in Networked Robotics, IISc, Bangalore, and Nokia CSR funds for their support.  
\bibliographystyle{elsarticle-num} 
\bibliography{main_bib}
\end{document}